\documentclass[12pt]{article}
\usepackage{ssi}
\usepackage{times}
\usepackage{epsf}

\begin{document}

\title{New Results from Super-K and K2K}

\author{R. Jeffrey Wilkes\thanks{Supported by DOE Contract DE-FG03-96ER40956. 
 \vskip 0.5in 
\noindent
\copyright\ 2002 by R. Jeffrey Wilkes. }\\ 
Dept. of Physics, Box 351560 \\
University of Washington, Seattle, WA 98195 \\[0.4cm]
Representing the Super-Kamiokande and K2K Collaborations
}

\maketitle
\begin{abstract}
This paper summarizes recent (as of {\it SSI-02}, in some cases 
updated in November, 2002) results from the 
Super-Kamiokande  and K2K experiments. 
The interpretation of Super-Kamiokande  
results on atmospheric and solar neutrinos provides strong
evidence for neutrino oscillations, hence non-zero neutrino mass.
While statistics are still limited, 
K2K data are consistent with Super-Kamiokande 
results on neutrino oscillations. The effort to
reconstruct Super-Kamiokande following a cascade of phototube implosions in 
November, 2001 is described. 
Plans for future are also discussed.
\baselineskip 16pt 
\end{abstract}

\section{Introduction}

This paper summarizes recent results from both Super-Kamiokande (SK) and
the KEK to Kamiokande (K2K) long baseline experiment, KEK Experiment E362,
which uses SK as its far detector.  SK and K2K
are independent experiments, operated by separate collaborations, although
there is some overlap in membership (including the present author).

  Both
of these experiments focus on the study of neutrino oscillations among
other topics.  While the Standard Model of electroweak interactions has
been tested with exceptional precision, it does not address the question
of the origin of generations and their mixing. Neutrino oscillations imply
that neutrinos are massive and that lepton flavors are not conserved
quantum numbers. Experiments on atmospheric neutrinos (including SK) 
found a significant path-length dependent deficit in the flux of
$\nu_{\mu}$ \cite{deficitALL:ref,subGeV:ref,multiGeV:ref}.  Experiments on
solar neutrinos (again, with SK prominently included) have similarly found
evidence for an apparent deficit in the flux of $\nu_{e}$
\cite{solar-review}.  The interpretation of these results provides strong
evidence for $\nu_{\mu}\rightarrow \nu_{\tau}$ and $\nu_{\mu}\rightarrow
\nu_{e}$  oscillations\cite{SKosc:ref,SKsterile:ref}. 

 In the two flavor approximation, which as we will see is reasonable for
SK and K2K data, for neutrino energy $E_{\nu}\rm (GeV)$ and path length
$L\rm (km)$ from production point to detector, the oscillation probability
can be written in terms of the mixing angle $\theta$ and the difference of
the neutrino masses squared $\Delta m^2$ (eV$^2$) as 
$$P(\nu_{\mu}\rightarrow\nu_x)=\sin^22\theta\sin^2 (1.27\Delta m^2
L/E_{\nu})$$
For SK, the path length varies from about 20~km for
neutrinos produced directly overhead, to about 13,000~km for upward going
neutrinos, which have passed through the earth, while K2K has a fixed path
length of 250~km.

In addition to the long-baseline analysis in K2K, the abundant neutrino
interaction sample obtained in the near detector allows measurement of
cross sections and other relevant data with better precision than many
previous published results. Thus the near detector in K2K provides validation
data for the Monte Carlo algorithm used to simulate neutrino interactions
in SK and K2K.

\section{Super-Kamiokande}

\begin{figure}[htb]
\begin{center}
\mbox{
	\epsfxsize=5.6truein
	\epsfbox{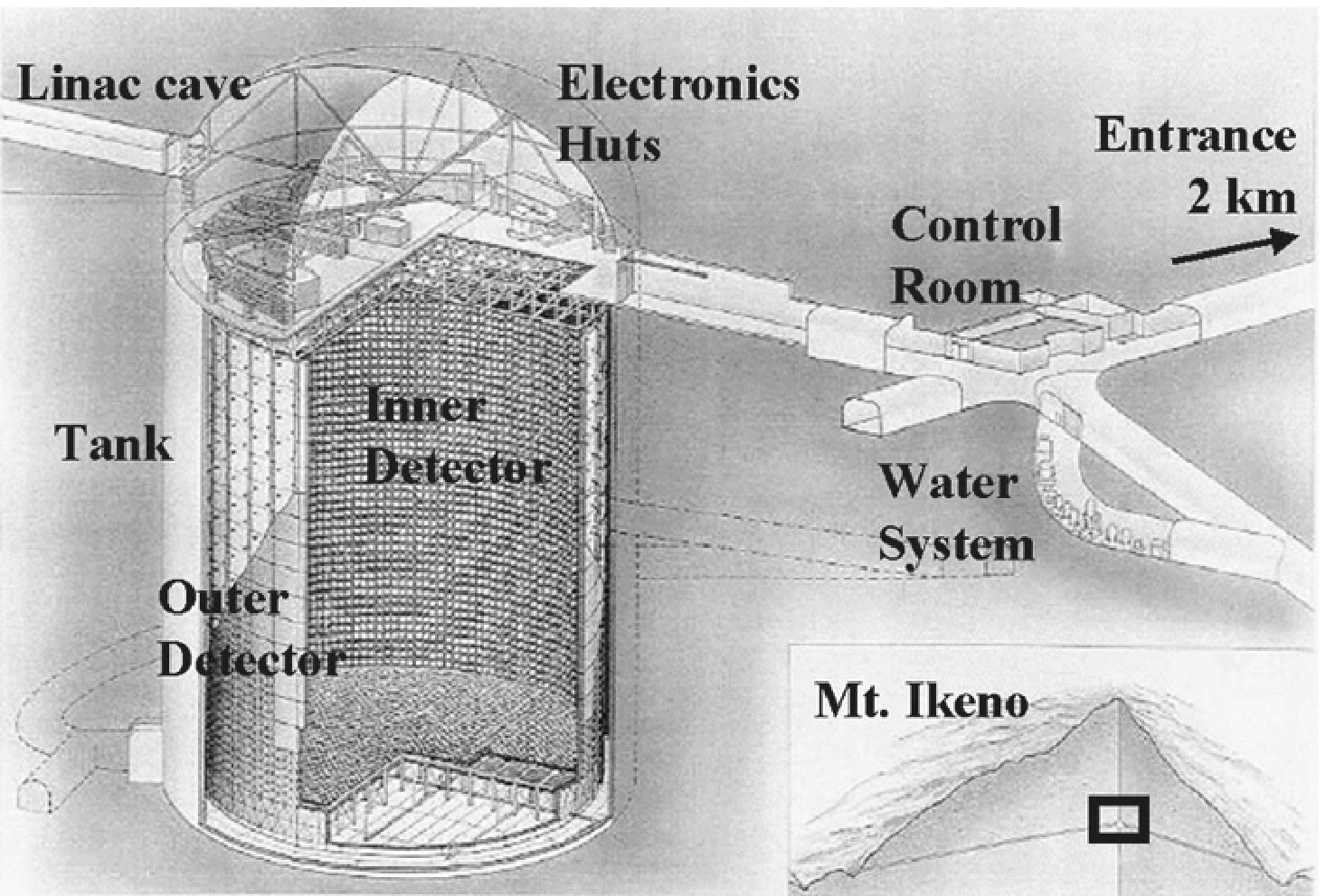}
}
\end{center}
\caption{The Super-Kamiokande underground water Cherenkov detector, located
near Higashi-Mozumi, Gifu Prefecture, Japan. Access is via a 2 km long truck 
tunnel. Inset shows location of detector area within the overburden.}
\label{fig:SK}
\end{figure}

SK is a water Cherenkov detector in the form of a cylinder with diameter
and height both approximately 40~m, and total contained mass 50~ktons
(Figure~\ref{fig:SK}).  The detector, which was commissioned in early
1996, is located $\sim$1000~m or 2700~m.w.e. (meters water equivalent) 
underground at the Kamioka Observatory, which is operated by the Institute
for Cosmic Ray Research of the University of Tokyo. The overburden is
sufficient to reduce the abundant surface flux of downward-going muons to
a level readily managed by the data acquisition system.  The detector is
divided by an optical barrier into a cylindrical primary detector region
(the Inner Detector, or ``ID'') and a surrounding shell of water
approximately 2.5~m thick (the Outer Detector, or ``OD'') serving as a
cosmic-ray veto counter. Details of the detector and general data
reduction procedures can be found in previous publications~\cite{subGeV:ref}. 

The OD is instrumented with over 1800 20-cm diameter Hamamatsu
photomultiplier tubes (PMTs). The ID was originally lined with
over 11,000 inward-facing 50-cm PMTs, providing 40\% photocathode coverage to
efficiently detect few-MeV solar neutrinos. In November, 2001, about half
of the 50-cm ID PMTs were destroyed in a cascade of implosions. Steps
taken to rebuild and restart the detector will be described below. Data
discussed in this paper were all taken during the period May,~1996 to
July,~2001, with the original (SK-I) detector configuration. This period
covered about 1489 live days of data-taking. 

Events in SK are categorized as due to solar neutrinos (visible energy
4$\sim$20 MeV) or atmospheric neutrinos (visible energy above 100 MeV). This
paper will focus on the atmospheric neutrino results; analyses of solar neutrino 
data have been published in detail elsewhere\cite{sk-solar02}. 
Atmospheric neutrino
events are further divided into fully-contained (FC, with no significant
activity in the OD), and partially-contained events (PC, with OD data
consistent with exiting particles). Fully-contained atmospheric neutrino
events are observed at an average rate of about 8.2 events per day, with
about 0.6 PC events per day.

Contained events are tagged as muon-like or electron-like by a likelihood
algorithm involving Cherenkov ring features, which was validated by a
beam test at KEK before SK data taking began\cite{KEK-beamtest}. In simple
terms, for the energy range of importance in SK, radiative losses are unimportant on average for muons, but dominant for electrons, resulting in significant showering (scattering and pair production). Thus muons leave sharp Cherenkov rings, while electrons leave fuzzier rings. 

Another atmospheric neutrino event category is 
upward going muons (upmus), either stopping (OD
data consistent with entering but no exiting particles), or through-going. 
The total cosmic-ray muon rate at Super-K is 2.2~Hz, of which a few
percent are stopping muons, and the overwhelming majority are
downward-going. Upmus can only be produced by neutrinos interacting in the
rock below SK, and on average represent 
the highest-energy sample of neutrinos,
but must be carefully separated from a substantial background of
near-horizontal downward-going muons.  The trigger efficiency for a muon
entering the ID with momentum more than 200~MeV/c is $\sim$100\% for all
zenith angles.  Stopping muons with track length \(>\) 7~m ($\sim$1.6~GeV) 
in the ID are selected for further analysis. 
Upmus are observed at the rate of about 1.4 events
per live day, of which about 0.3 per day are stopping tracks.

Details of the track reconstruction method and data reduction and analysis
algorithms used in SK-I have been described in earlier papers~\cite{SKupmu}.

\section{Results from SK-I atmospheric neutrino data}

\begin{figure}[htb]
\begin{center}
\mbox{
	\epsfxsize=5.6truein
	\epsfbox{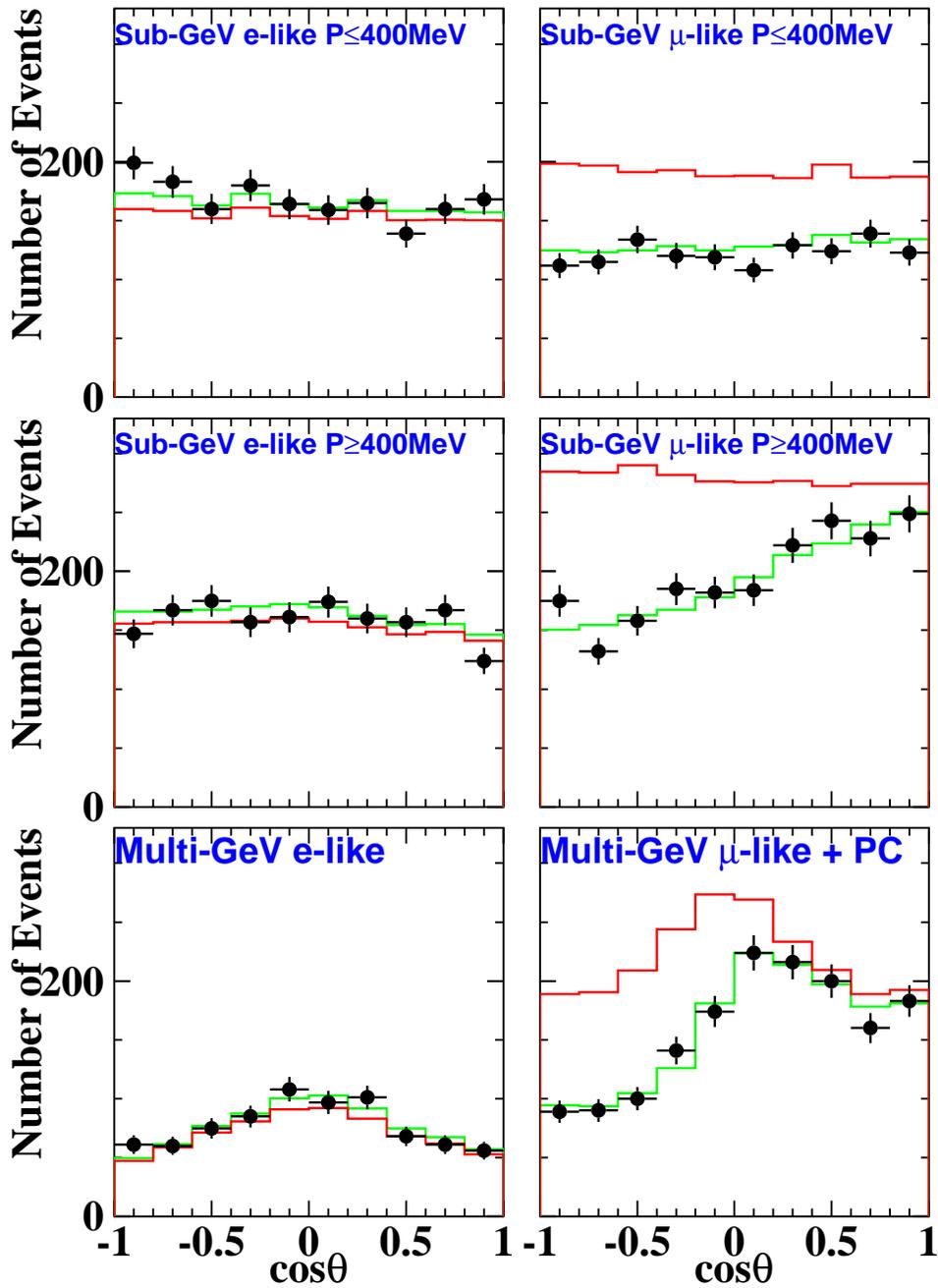}
}
\end{center}
\caption{Atmospheric neutrino zenith angle distributions in SK (data points). 
Electron-like events 
agree with expectation for no oscillations (darker histogram), 
but muon-like events show a large deficit
for upward-going neutrinos (i.e., those with long path lengths). Lighter 
histogram shows expectation for best-fit oscillation parameters.}
\label{fig:sk-angdis}
\end{figure}

SK observes a significant deficit in muon-like atmospheric neutrino
events, relative to expectation based on the best available simulations of
cosmic ray interactions, and the subsequent production and propagation of
neutrinos through the atmosphere. The deficit is strongly zenith-angle
dependent, with a large effect for upward-going muon neutrinos and little
or no deficit for downward going neutrinos of either flavor
(Figure~\ref{fig:sk-angdis}). At the energies considered, absorption in
the earth is unimportant (and taken into account in the analysis). The
zenith angle of arrival is of course correlated directly with the
neutrino's flight path from its point of origin in the atmosphere. 
Downward-going neutrinos have travelled on the order of 20~km or less,
while upward-going neutrinos have travelled up to 13,000~km through the
earth. Thus neutrino oscillation is the leading suspect to explain the
deficit. 

\begin{figure}[htb]
\begin{center}
\mbox{
	\epsfxsize=5.6truein
	\epsfbox{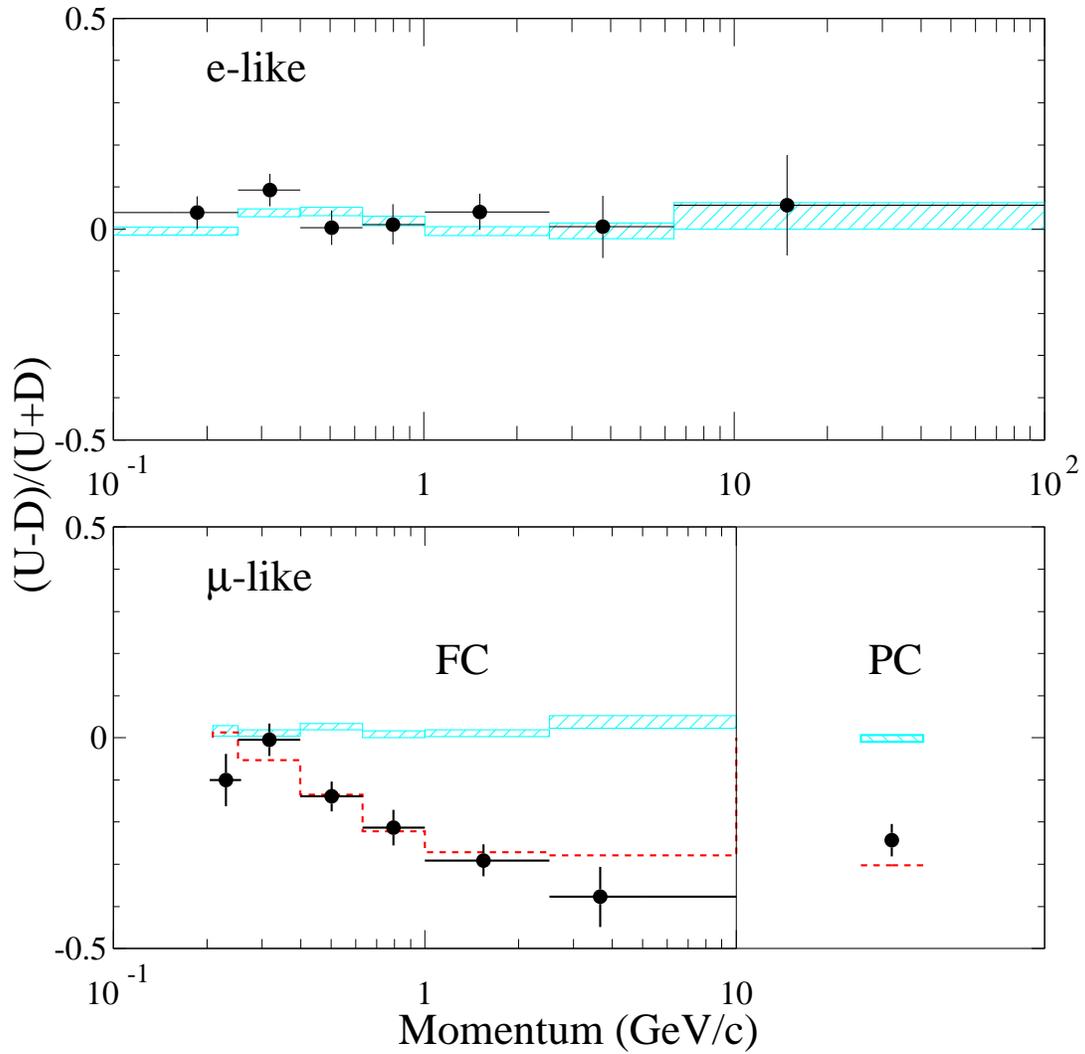}
}
\end{center}
\caption{Up/Down ratio for atmospheric neutrino events in SK, an integral
representation of the angular distributions. A 
highly significant deficit 
relative to expectation for no-oscillations is seen for muon-like events
while electron-like events are consistent with expectation. Dashed 
histogram in lower plot shows expectation for best-fit oscillation 
parameters.} \label{fig:sk-ud}
\end{figure}

Both the angular distributions of various event categories and the up/down
ratio (Figure~\ref{fig:sk-ud}), which provides a relatively
high-statistics integral measure of the effect, deviate from the
no-oscillations expectation by many standard deviations, while agreeing
well with internally consistent sets of oscillation parameters. Since
electron neutrino distributions show no significant path-length dependence, 
it appears
that SK observes only muon neutrino disappearance. The size and
configuration of SK make it inefficient for identifying tau neutrinos,
which promptly decay to muon and electron neutrinos. Muon neutrino
disappearance, with observed electron neutrino rates approximately
according to expectation, suggests that oscillations in the region of
energy and path length probed by SK are dominated by 2-flavor $\nu_\mu-\nu_{tau}$
oscillations. 

\begin{figure}[htb]
\begin{center}
\mbox{
	\epsfxsize=5.6truein
	\epsfbox{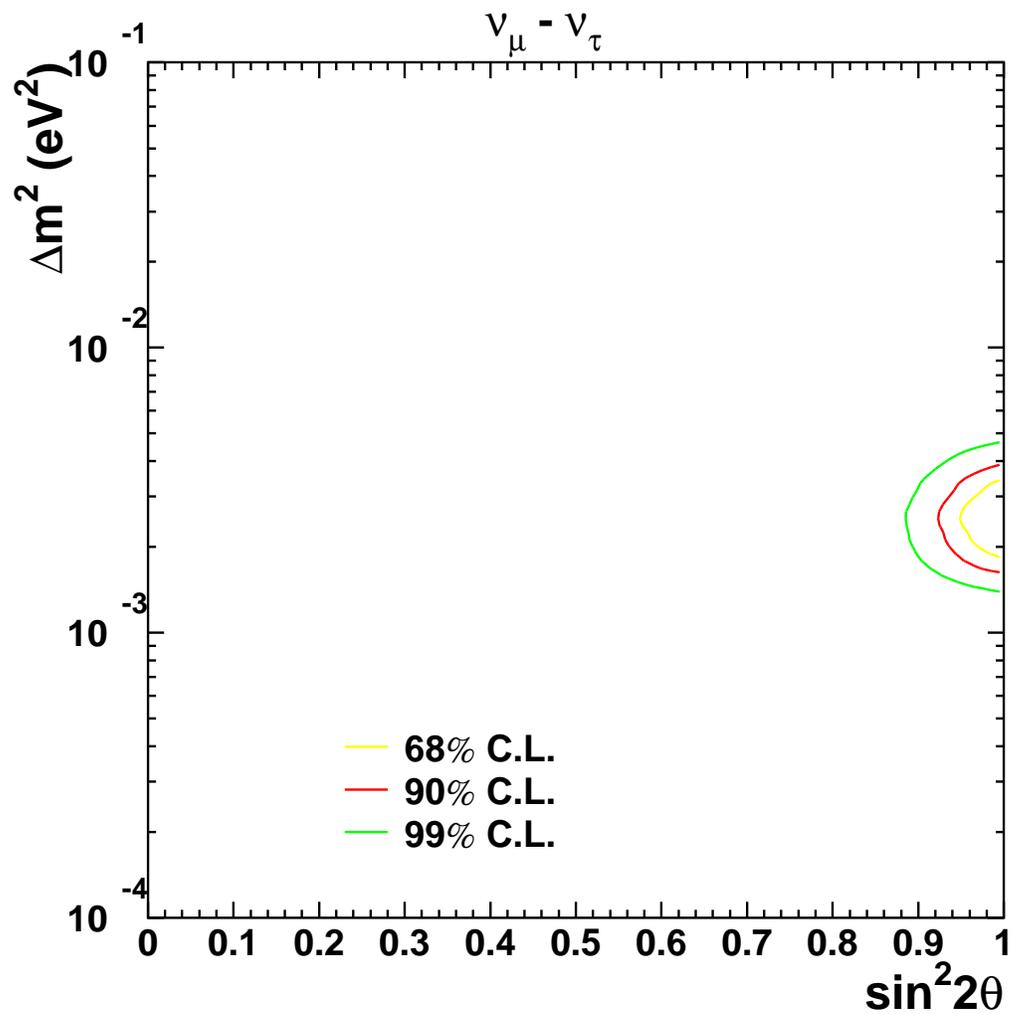}
}
\end{center}
\caption{Allowed region in 2-flavor oscillation parameter space from combined
analysis of FC, PC and upmu events in SK. 
The fit shown was constrained to occupy only the
physical region ($\sin^2 2\theta \leq 1.0$); a fit without this constraint 
gives very similar results within the physical region. }
\label{fig:combined-allowed}
\end{figure}

FC, PC and upmu data were reduced and analyzed by independent procedures
and the fact that they give consistent results on best-fit oscillation
parameters is significant. For the combined data sets, the observed number
of events, binned in energy and zenith angle, are compared to the
simulation results, which assume no oscillations, and include allowances
for systematic errors on the atmospheric neutrino production and
propagation algorithms, as well as incorporating a large variety of
calibration data and detector response measurements. The Monte Carlo (MC) 
results thus cover all known detector effects and efficiencies, while
systematic error allowances for factors unamenable to calibration or
direct correction are included as free
parameters in the fitting procedure. A chi-squared minimization is used to
find the allowed-region boundaries for selected confidence levels, on the
2-dimensional oscillation parameter space ${\Delta m^2, \sin^2(2\theta)}$. 
The latest released results for the combined analysis, 
covering the full SK-I period,
are shown in
Figure~\ref{fig:combined-allowed}. As the plot shows, the atmospheric neutrino data suggest $\{\Delta m^2, \sin^2(2\theta)\}=\{1.5\sim 4\times 10^{-3}{\rm eV}^2 , \sim 1.0\}$, {\it i.e.}, full mixing.

While cosmological and big-bang nucleosynthesis considerations 
limit the number of light active neutrinos to about four\cite{cosmo-N-nu}, 
LEP experiments\cite{LEP-N-nu} constrain the number of active
neutrino flavors to $2.99\pm 0.01$. Combined results from solar neutrino
experiments, atmospheric neutrino experiments, and nuclear reactor
neutrino oscillation experiments can accomodate a 3-flavor mixing scheme. 
However, in order to simultaneously accomodate results from the LSND
short-baseline experiment\cite{LSND}, it is necessary to assume a fourth
type of neutrino, which must therefore be ``sterile": unable to interact
with matter. Super-Kamiokande analyzed its atmospheric neutrino data to
see if limits could be placed on the existence of sterile neutrinos, and
found that they could be ruled out at the 99\% confidence level, as
described in detail elsewhere\cite{SKsterile:ref}. 

\section{K2K}

Long-baseline experiments, in which a neutrino beam is directed from a
particle accelerator laboratory to a distant, off-site detector, were
first seriously proposed in the 1970s\cite{mann-primakoff}, but 
K2K was the first
such experiment to actually be commissioned and take
data\cite{k2k-paper01}. Several new long-baseline experiments are now
under construction or in advanced planning stages\cite{minos,cern-lbl},
including a next-generation version of K2K\cite{JHF2K}.

\begin{figure}[htb]
\begin{center}
\mbox{
	\epsfxsize=5truein
	\epsfbox{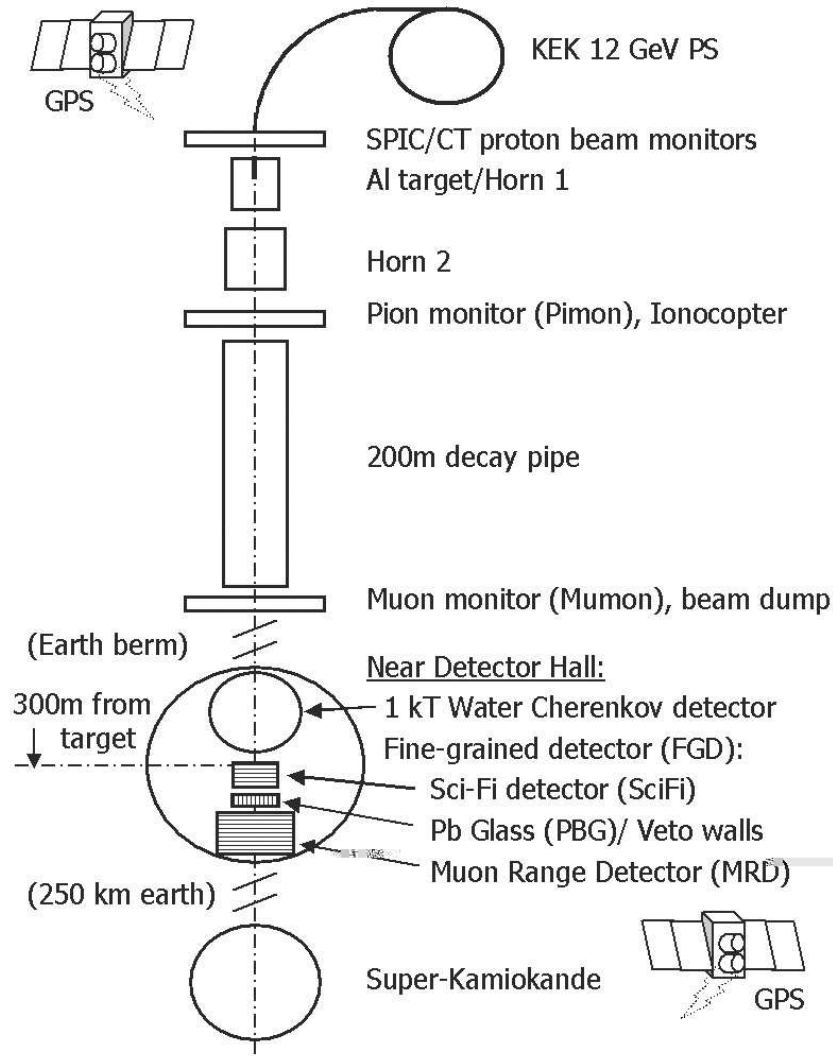}
}
\end{center}
\caption{Overview of K2K, showing cartoon view of important elements: beamline, 
beam monitors, near detector, far detector, and GPS time synchronization system. 
Beam spill times 
are timestamped at KEK using a GPS clock system. The list of spill times 
is transmitted by Internet link to SK, and used to select events,
which are timestamped by an identical GPS clock system, 
before they leave the offline processing stream. SK triggers within a 
1.5 $\mu$sec acceptance window
around the expected arrival time of beam neutrinos are analyzed as K2K events.}
\label{fig:k2k-concept}
\end{figure}

Figure~\ref{fig:k2k-concept} shows the conceptual design of K2K. The KEK wideband
neutrino beam is generated by directing a
proton beam from the 12 GeV KEK Proton Synchrotron (PS) toward SK,
including a $1^\circ$ dip angle into the
 earth.  Every 2.2 s, approximately $6\times 10^{12}$ protons in nine
bunches are fast-extracted, making a 1.1 $\rm\mu{s}$ beam spill. 
Following a 3-cm diameter Al target embedded in the front element of a
2-horn magnet system, positive pions are focused into (and negative pions
diverted from) a 200~m long decay pipe. A pion monitor (PIMON) employing a
novel gas Cherenkov detector system
 can be inserted in the beamline to sample the secondary pion
distributions\cite{PIMON}. Pion decays yield primarily muon neutrinos and
muons. The muons are sampled by a muon monitor detector (MUMON) at the end of the
decay tunnel, and then stopped by an iron and
 concrete beam dump and 70 m of earth, leaving a beam of muon neutrinos 
(98\% pure) to make its way through the earth's crust to SK. 

K2K originally used a 2~cm diameter Al target, for which the acceptance of
the horns and decay pipe were designed. However, it was necessary to
increase the target diameter after initial test runs revealed 
problems with the original design. This change reduced the efficiency of
the horn system, lowering the net expected neutrino flux at SK for the
integrated KEK beam allocation of $10^{20}$ pot. Thus the total number of
neutrino events expected in SK for the full exposure (in the absence of
oscillations) will be about 200. After the latest run, which ended in
July, 2001, the accumulated exposure was 5.6$\times10^{19}$ pot, or about
1/2 the expected total for the experiment. Possible extensions of the K2K
program will be mentioned below.

\begin{figure}[htb]
\begin{center}
\mbox{
	\epsfxsize=5.6truein
	\epsfbox{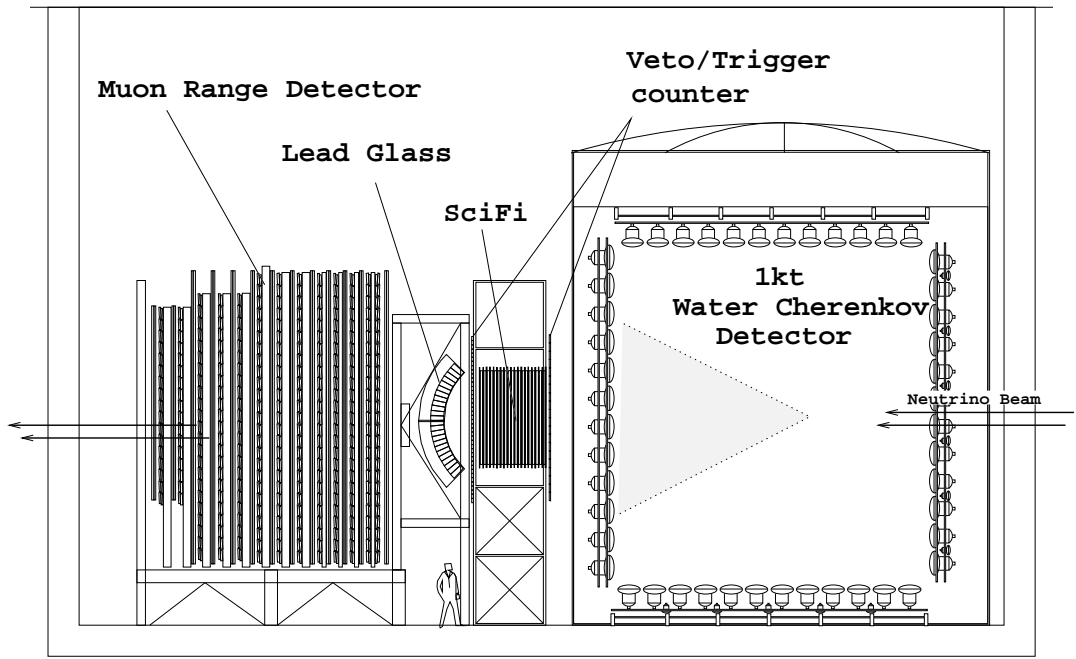}
}
\end{center}
\caption{Near detector components at KEK. The 1kt Water Cherenkov detector employs identical
PMTs and geometry as SK. The SciFi detector, with water target tanks, provides detailed
tracking for analysis of quasi-elastic and inelastic events. 
The Lead Glass detector was removed in 2001 and will be replaced with 
another tracking detector\cite{scibar}.
The Muon Range detector identifies
and measures the momentum of muons; accumulated data provide information 
on beam stability.}
\label{fig:k2k-near}
\end{figure}

The near detector suite (Figure~\ref{fig:k2k-near}) is located 
in an underground experimental hall
300~m from the production target, where neutrino interactions are observed
by a set of detectors with complementary capabilities.

The K2K one kiloton water Cherenkov detector (1kt) uses the same
technology and analysis algorithms as SK, the far detector.  It has 680
50-cm photomultiplier tubes (PMTs) on a 70-cm grid lining 
a $\rm 8.6 m$
diameter, $\rm 8.6 m$ high cylinder.  The PMTs and their arrangement are
the same as in SK, and give the same fractional photocathode coverage
(40\%) as in SK-I.  A scintillating fiber detector
(SciFi)\cite{SciFi-NIMpaper}, consisting of 6 tons of target (water in
thin Al tanks) interleaved with sci-fi tracking layers, allows
discrimination between different types of interactions such as
quasi-elastic or inelastic.  Downstream of the SciFi there was a lead
glass array for tagging electromagnetic
showers, which was removed after the 2001 run. It will be replaced with 
a scintillator-bar array\cite{scibar} (SciBar) to be installed during the summer 2003 
beam shutdown. 
A muon range detector (MRD)\cite{MUC:ref}, of total mass 915
tons, measures the energy, angle, and production point of muons from
charged current (CC) $\nu_{\mu}$ interactions.

K2K beam-induced events in SK are selected by comparing SK high energy event 
trigger times
with the KEK-PS spill times via timestamps generated at
both sites using identical Global Positioning System
(GPS) clock systems\cite{GPS:ref}. Absolute times at the near and far sites are
synchronized within about 100 nsec. The rate of background events due to
atmospheric neutrinos in the timing acceptance window of 1.5 microsec
every 2.2 sec is negligible: $\sim10^{-3}$ events for the whole
experiment.

      The beamline was aligned by a GPS position survey\cite{align:ref}. 
The precision of this survey for defining the line from target to far
detector is better than 0.01 milliradian, and the precision of
construction for the near site beamline alignment is better than 0.1 mr. 
However, only about $\rm 3 mr$ accuracy is required for beam pointing,
since the predicted neutrino spectrum at 250 km is effectively constant
over a diameter of nearly 1 km.

MUMON data are used to monitor the steering of the $\nu_{\mu}$ beam in
realtime.  The muon yield is directly correlated with the horn-focused
$\nu_{\mu}$ beam intensity. MRD data provide a long-term stability check.
The MRD event rate normalized to measured proton beam intensity is
constant within errors, implying the $\nu_\mu$ energy spectrum was stable
throughout the K2K run period.

      To predict the $\nu_{\mu}$ beam characteristics at the far site, a
normalization measurement at the near site and extrapolation from near to
far are necessary.  For event rate normalization, only the 1kt data are
used, to take advantage of the cancellation of 
detector and analysis systematics in the
far/near ratio. The average $\nu_\mu$ event rate per proton on target
(pot) in the 1kt detector is $3.2\times 10^{-15}$.

 For the near to far extrapolation, a beam Monte Carlo 
simulation is used, which is validated
using the PIMON measurements of secondary pion momentum and angular
distributions.  This simulation is based on GEANT\cite{GEANT:ref}, with
detailed descriptions of materials and magnetic fields in the target
region and decay volume.  It uses as input a measurement of the primary
beam profile at the target.  Primary proton interactions on aluminum are
modeled with a parameterization of hadron production data\cite{hadr:ref}.
Other hadronic interactions are treated by GEANT-CALOR\cite{GCALOR:ref}.

\begin{figure}[htb]
\begin{center}
\mbox{
	\epsfxsize=5.6truein
	\epsfbox{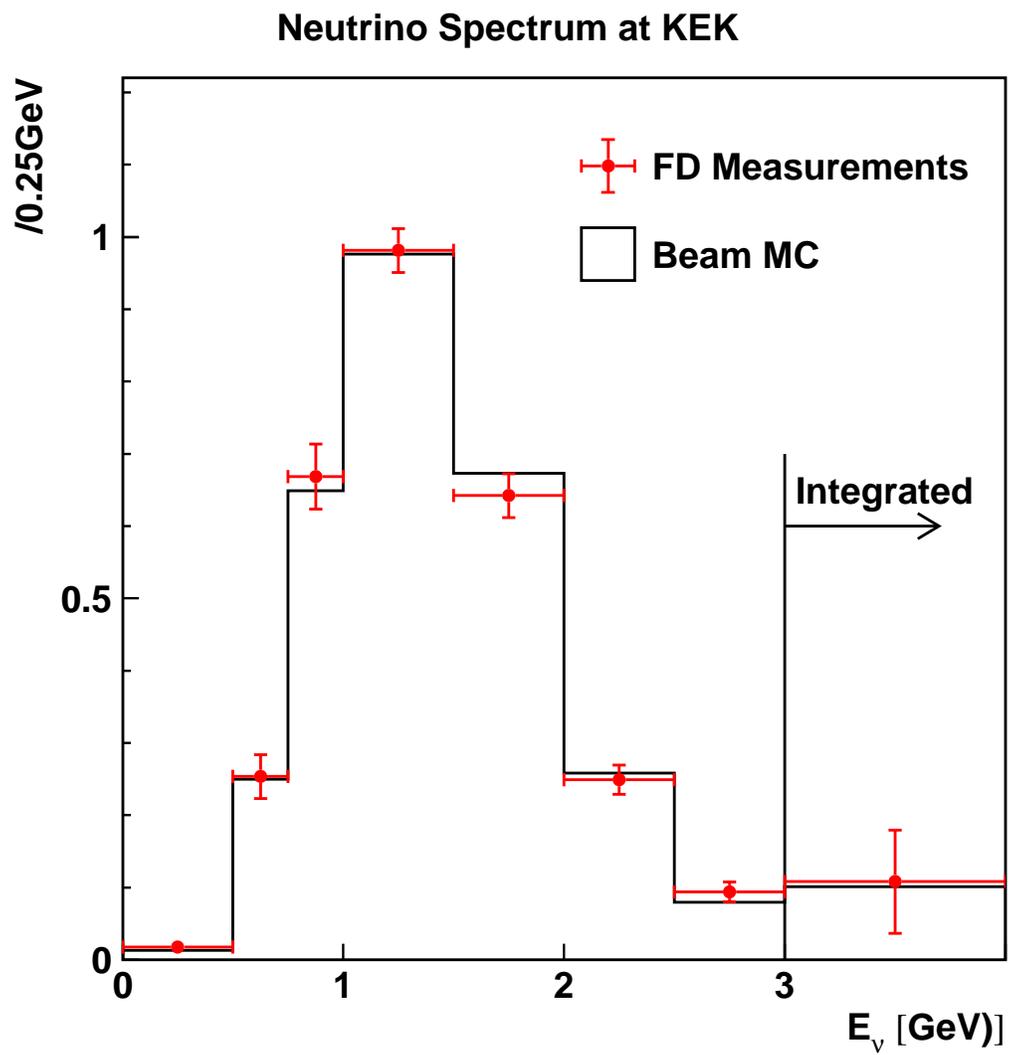}
}
\end{center}
\caption{K2K neutrino beam energy spectrum, as inferred from front (near)
 detector data, 
compared to the beam Monte Carlo. }
\label{fig:beammon}
\end{figure}

Once the kinematic distribution of pions after production and focusing is
known, it is possible to predict the $\nu_{\mu}$ spectrum at any distance
from the source.  Correcting for efficiencies, relative target masses, and
detector live times, the expected $\nu_\mu$ signal at the far site can be
estimated.  Figure~\ref{fig:beammon} shows the inferred $\nu_\mu$ energy
spectral shape at the near site 
along with the beam simulation result.  As can be seen, the beam
simulation is in excellent agreement with near detector data.

Data reduction algorithms similar to those used in atmospheric neutrino
analyses at SK\cite{subGeV:ref,multiGeV:ref} are applied to select fully
contained (FC) neutrino interactions. The number of Cherenkov rings and
particle identification are reconstructed by the same algorithms as those
used at SK\cite{subGeV:ref}, and the event category definitions are also
the same.

\begin{figure}[htb]
\begin{center}
\mbox{
	\epsfxsize=5.6truein
	\epsfbox{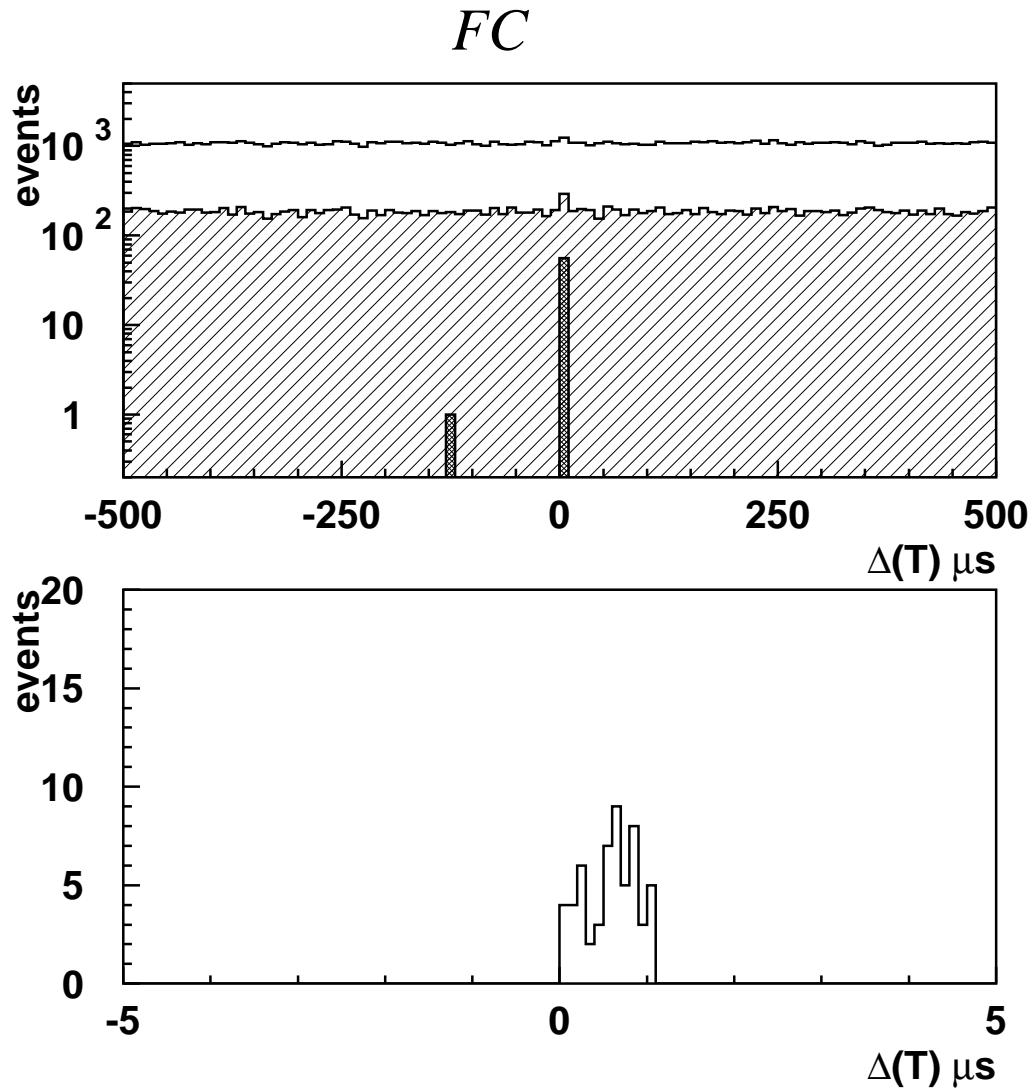}
}
\end{center}
\caption{Time difference between triggers at SK and expected time of beam neutrino arrival,
based on spill time data from KEK. 
Upper histograms show various stages in the data reduction;
after the required fiducial volume and energy cuts are imposed, 
it can be seen that a sharp peak appears at 0, relative to the 
$1.5\mu$sec timing acceptance window. The nearest random background event 
(out of the entire data sample) can be seen over $100\mu$sec out of time. Lower plot
shows a closeup of the region around $\Delta{T}=0$.}
\label{fig:tdiff}
\end{figure}

Since the measured uncertainty of the time synchronization accuracy for
the two sites is $\rm<200 ns,$ beam-induced events are selected by
defining a $\rm 1.5 \mu{s}$ acceptance window surrounding the $\rm 1.3
\mu{s}$-long spill time every 2.2 sec, offset by light-speed transit time to SK. 
Figure~\ref{fig:tdiff} shows the $\Delta(T)$ distribution at successive
stages in the SK data reduction. After various cuts used to select high
energy events, a clear, isolated peak in time appears, correlated with the
predicted arrival of the neutrino beam from the KEK-PS.

The K2K oscillation analysis uses two independent approaches, which can
then be combined in a global likelihood fit for oscillations parameters. 
First, the total number of beam-induced events in SK is compared to the
expected number of events in the absence of oscillations, taking into
account the coordinated livetimes of SK and the various near detectors,
and the corresponding integrated beam intensity (as measured by the number
of protons on target delivered during simultaneous detector livetimes). 
This provides a
relatively high statistics integral measure of oscillation effects, and
allows a test of the no-oscillation hypothesis with maximum significance
given the limited data. 

Second, the shape of the neutrino energy spectrum is a sensitive indicator
of oscillation effects. Since path length $L$ is fixed in K2K, any dependence
on $L/E$ reveals itself in distortions of the neutrino energy
spectrum. 

\begin{figure}[htb]
\begin{center}
\mbox{
	\epsfxsize=5.6truein
	\epsfbox{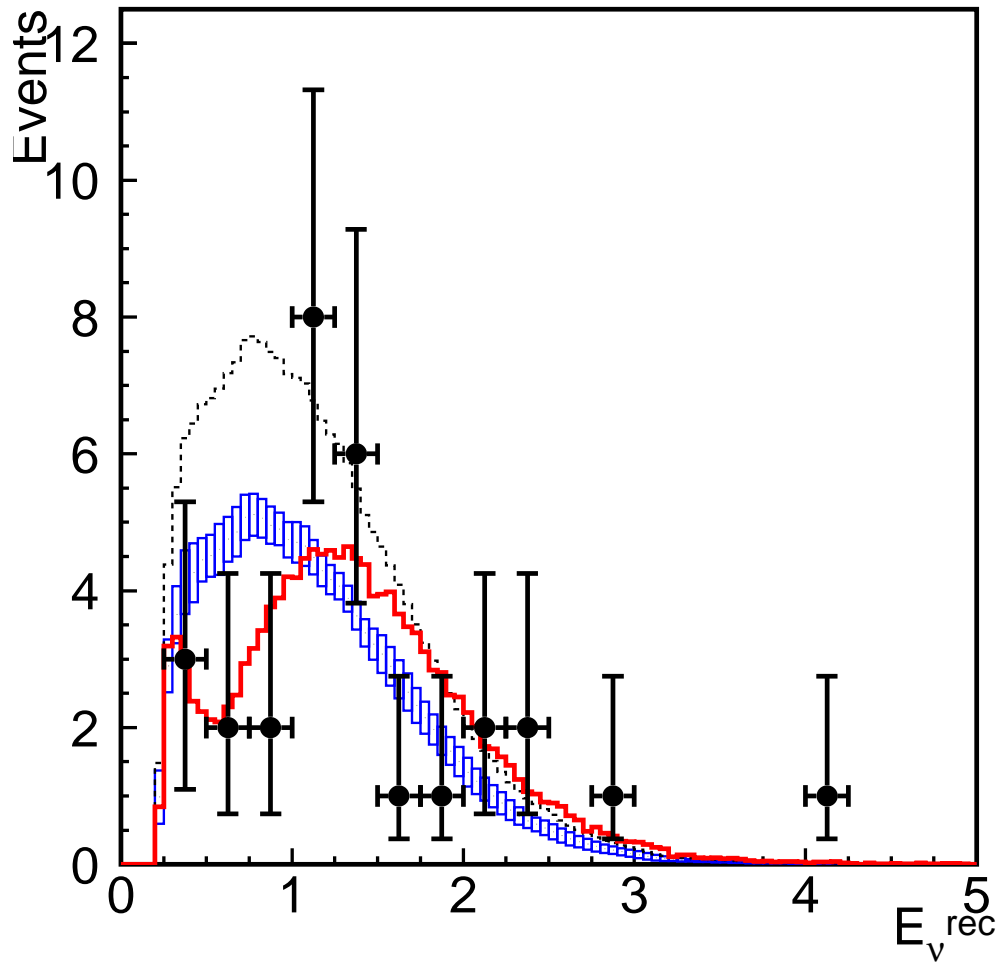}
}
\end{center}
\caption{Energy spectrum of the K2K events at SK. Predicted spectrum
obtained by extrapolation of near detector results is compared with 
observed SK data. Points with error bars are data. Boxes:
expectation for no oscillations, showing systematic errors, normalized to
the number of observed events.  Solid: spectrum for 
best-fit oscillation parameters, also normalized to the number of observed 
events. Dashed line: 
expectation for no oscillations, normalized to {\it expected} number of events.}
\label{fig:k2k-spec}
\end{figure}

\begin{figure}[htb]
\begin{center}
\mbox{
	\epsfxsize=5.6truein
	\epsfbox{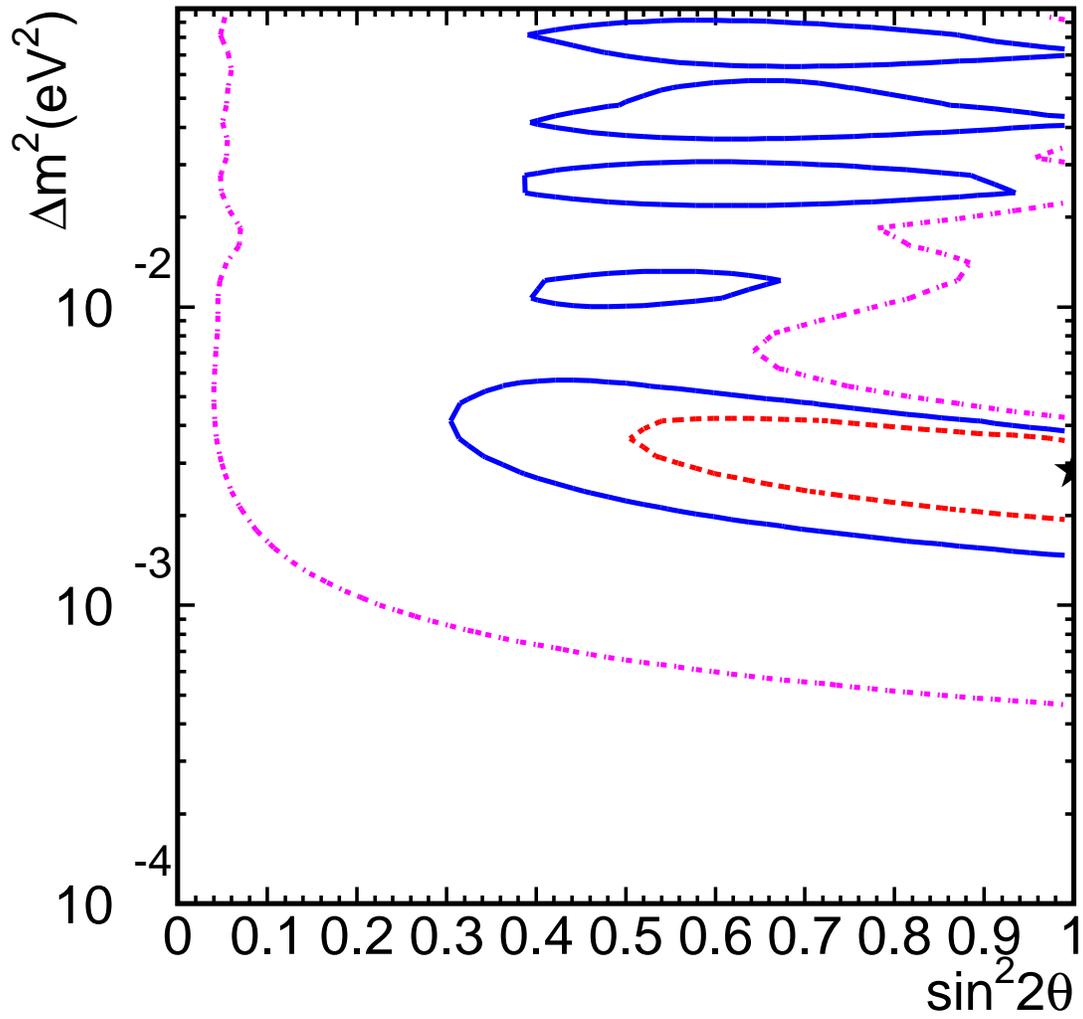}
}
\end{center}
\caption{Allowed region in 2-flavor ($\nu_{\mu}\rightarrow \nu_{\tau}$) 
parameter space from 
K2K combined analysis of the number of events and 
energy spectrum data. Solid line indicates 95\% CL contours; 
dashed line: 68\%; dot-dashed line: 99\%; star indicates best-fit point. Compare to 
Figure~\ref{fig:combined-allowed}.}
\label{fig:k2k-allowed}
\end{figure}

The predicted number of SK events in the absence of oscillations, 
for the running period discussed here,
is $80.1^{+6.2}_{-5.4}$. With 56 observed
FC events in SK, the probability that the observed number is due to chance
fluctuations is less than 1.3\%.  
Figure~\ref{fig:k2k-spec} compares the 
observed far detector energy 
spectrum in comparison with expectation for the no oscillations hypothesis, 
and the best-fit oscillation parameters. 
Putting together 
the spectrum shape and number information in a combined fit, the
no-oscillations hypothesis can be rejected with 99.3\% confidence. 
Figure~\ref{fig:k2k-allowed} shows the allowed region in parameter space; 
these results indicate $\Delta m^2=1.5\sim 4\times 10{-3}{\rm eV}^2$, with large mixing angle, quite consistent with the SK allowed region.

While statistics are still limited, K2K data collected during the SK-I run are 
evidently consistent with SK results on atmospheric neutrino oscillations. 
A paper
submitted at time of writing\cite{k2k-paper02} describes in more detail
the methods used to obtain these results.
Additional data will be collected starting in January, 2003, and another run is 
planned for 2004. Further data taking may be possible
depending upon the interplay between 
KEK-PS operations and construction of the new 50 GeV accelerator
at the Japan Atomic
Energy Research Institute (JAERI) site, as described below.

\section{Deconstruction and Reconstruction of Super-K}

As noted, a cascade of PMT implosions in November, 2001 destroyed over
half of the PMTs in the SK detector. Thanks to a prompt response by
agencies supporting the experiment (and intense effort by many people), it
was possible to begin reconstruction a few months thereafter, with an
aggressive timetable calling for restarting the experiment in December, 2002. 
By the end of 2001, an investigation had been conducted, causes
identified, and a plan for protecting against future occurrences developed
and tested. 

SK-II ID PMTs are contained in plastic housings, with a clear acrylic dome
over the photocathode, and the tube body and base enclosed in a PVC shell. 
The acrylic domes have a negligible (few percent) effect on photon
detection efficiency, since they are only a few mm thick and their index
of refraction is close to that of water. The housings are not pressure
vessels, but are simply intended to slow down the influx of water in case
of implosion, eliminating the shock wave effect that initiated the cascade
of
 implosions in late 2001. Destructive testing was performed at 30~m water
depth, with PMTs closely spaced as in SK-I, both with and without
housings, and performance of the protection scheme was confirmed. Tests
showed that implosion of a 50-cm PMT will not harm its nearest neighbors
with the housings in place. Of course, during SK-II the approximately
doubled spacing between ID PMTs will provide an additional margin of
safety. 

Since it was not possible to procure a full complement of 50-cm PMTs
promptly, the resurrection of SK will proceed in two stages. SK-II,
scheduled to begin data taking in December, 2002, will have approximately
50\% ID PMT coverage relative to SK-I. Surviving PMTs and about 1000 new
PMTs were redistributed around the ID, with roughly every other PMT site
occupied. The reduction in photocathode coverage will raise the effective
threshold for solar neutrinos, but simulation studies indicate that data
from high energy triggers (atmospheric neutrinos, upmus and K2K beam
neutrinos) will suffer only minor reductions in angular and energy
resolution. By 2005, enough 50-cm PMTs to will be available to restore
full occupancy in the ID; this upgrade will initiate the second stage of
revival, SK-III.

Progress in carrying out these plans has been rapid and efficient. By the
end of March, 2002, debris had been cleared and reconstruction work begun. 
All PMTs were removed before starting reconstruction. Sufficient spares
were available to replace all lost
 20-cm OD PMTs, so the OD was restored fully. Repairs to the bottom and
side PMTs were completed by September, 2002, and filling with 
water began. The detector was operated throughout the filling period,
except for brief intervals for calibration
 studies, to maintain a supernova watch with the available volume. 
By the time of writing (November, 2002), the SK tank had been refilled
without incident, and final work on the top layer of PMTs was being
completed. Thus SK is on track to begin data taking by the time the next
K2K beam is available. Beam tuning at KEK will begin in mid-December, 2002. 

\section{Future} 
K2K will take data throughout 2003, except for the usual KEK-PS summer beam 
shutdown. By 2005, it is expected that K2K will have
received its allocated beam exposure, and KEK beamline components will 
need to be
moved to the new 50 GeV accelerator under construction at the JAERI site
 at Tokaimura, about
130~km NE of Tokyo. The new high-intensity accelerator, recently
officially renamed J-PARC (Japanese Proton Accelerator Research Complex) 
but still commonly referred to by its original acronym ``JHF" (Japan
Hadron Facility), will deliver first beam in 2007. By then, SK will have
been re-equipped with a full complement of PMTs. 

At that time, interest will be focused on a second-generation long-baseline
experiment, with working title ``JHF-Kamioka". Preliminary steps have been taken
to organize a broadly-based international collaboration. 
Initial designs for a narrow-band neutrino beamline,
 and conceptual designs for a near
detector configuration have been developed and are under discussion by
interested physicists\cite{JHF2K}. The JHF neutrino beam will deliver more events per
unit time to SK than the KEK-PS beam by about two orders of magnitude. 
Thus JHF-Kamiokande should be able to fully explore the
parameter space defined by SK and K2K with high statistics within a few
years. The baseline from Tokaimura to SK is 295~km, 
only slightly larger than for K2K.

The next goal for long-baseline neutrino physics will be to explore CP
violations, requiring much higher statistics. For this task, 
SK is too small to contribute significantly in reasonable time, 
even with the JHF beam. Preliminary plans, still at the
conceptual level, have been developed for
 ``Hyper-Kamiokande", a 1-megaton water Cherenkov detector\cite{hyperk}. 
The SK geometry
cannot simply be scaled up, due to fundamental engineering limits on cavity
size in rock, but a gallery of adjacent units could be constructed. 
Also, the present SK
site cannot be used, due to limitations on road access to the Mozumi mine. 
However, a site close to the town of Kamioka, where the mining
company has its main facilities, could be used. The neutrino beam will be 
designed to be broad enough to cover both sites with its stable central
region.

\section{Conclusion}
There is growing consensus that there are probably only three low-mass 
neutrinos (unless Mini-Boone confirms LSND results\cite{miniboone}), with 
the lighter pair of
eigenstates nearly degenerate and the third state separated in mass. 
As a result, we observe two mass-squared
differences, $\Delta m_{12}^2$ and $\Delta m_{23}^2$, 
with different magnitudes. The larger $\Delta m^2$  
can be identified with the two-state $\nu_\mu-\nu_\tau$ mixing observed
in atmospheric oscillations, while the smaller $\Delta m^2$ applies to
the $\nu_\mu-\nu_e$ mixing observed in solar neutrino
experiments such as SK and SNO.
The mass
hierarchy, {\it i.e.}, whether the third state lies above or below the close
pair, is yet to be determined. 
However, exploration of the neutrino sector is far from complete, and second
generation experiments such as MINOS, JHF2K, ICARUS, {\it et al} 
have much to do.
New experiments on mixing schemes, mass hierarchy, matter effects, 
and perhaps CP violations
will keep the field lively for some time to come. Interested students are
invited to join the fun!

\section{Acknowlegements}
The author gratefully acknowledges the cooperation of all other members of
the Super-Kamiokande and K2K collaborations, 
who produced the results discussed, and 
reviewed this paper. Any remaining
errors or misstatements are my responsibility alone. The full membership
of each collaboration can be found in their respective recent
publications\cite{sk-solar02,k2k-paper02}. 
The Super-Kamiokande experiment and the
K2K experiment have been built and operated with support from the Japanese
Ministry of Education, Culture, Sports, Science and Technology, and the
United States Department of Energy. The author gratefully acknowledges
support from DOE grant DE-FG03-96ER40956, KEK, 
and SLAC, and thanks the SLAC Summer
Institute staff for their hospitality.

\end{document}